\documentstyle[pre,aps,multicol,epsf]{revtex}

\begin{document}
\title{Bound spin-excitons in two-dimensional electron gas}
\author{V. Fleurov$^{1}$ and K. Kikoin$^{2}$}
\address{
$^1$School of Physics and Astronomy, Beverly and Raymond Sackler
Faculty of Exact Sciences.\\ Tel Aviv University, Tel Aviv 69978,
Israel.\\ $^2$Physics Faculty, Ben-Gurion University of the
Negev\\ Beer-Sheva 84105, Israel.}
\date{\today}
\maketitle
\begin{abstract}
A theory of the spin exciton capture by a magnetic impurity in a
2D electron gas is developed. We consider the resonance model for
electron scattering by a transition metal impurity and calculate
the binding potential for spin excitons. This potential is spin
selective and is capable of binding a spin exciton with zero
angular momentum. In order to trap an exciton with a nonzero
angular momentum $m$, the potential must exceed a certain
threshold value, depending on $m$.
\end{abstract}
\section{Introduction}
Shubnikov-de Haas effect is a powerful tool for studies of a
two-dimensional electron gas (2DEG) in a strong magnetic field. In
particular, the oscillatory behavior of the electronic $g$ factor
in 2DEG was investigated with the help of this effect (see
\cite{Nic} and references therein). Two important parameters which
predetermine the properties of the 2DEG in a strong magnetic field
$B$ are the cyclotron frequency $\omega_B=eB/mc$ and the effective
Coulomb energy $E_c = e^2/\kappa l_B$ $(l_b=(\hbar c/eB)^{1/2})$
is the magnetic length). In the limit of ultra high magnetic
fields, when $E_c\ll \hbar \omega_B$ and only the lowest Landau
sublevel is filled, the low-energy branches of the excitation
spectrum are represented by well separated bands of spin waves,
magnetoplasmons, etc. These excitations were studied in details
during recent decades (see, e.g.,
\cite{Lelo80,Bychok81,KH84,KH85,AM91,LK93}). The spin waves,
formed by the electrons in the down-spin and holes in up-spin
$n=0$ Landau subbands, make the lowest branch of magnetic
excitations. These states are separated by the Zeeman gap
$\Delta_B = g\mu_B B$ from the ground state and form a band with
the width of $\sim E_c$. The exciton dispersion law is quadratic
at small wave numbers and saturates in a short-wave limit, where
the excitons, in fact, transform into free electron-hole pairs
\cite{Lelo80}.

This paper studies the interaction between the spin waves and a
magnetic impurity in 2DEG. To be more specific, we consider doped
heterojunctions GaAs/GaAlAs and related materials, so we refer
below to a III-V semiconductor as a host material where some
cation atoms are substituted for magnetic transition metal
impurities. It is known \cite{KF94}, that transition metal atoms
create deep levels in the forbidden energy gap of the host
semiconductor, and the main mechanism of the electron-impurity
scattering is the resonance scattering by the $d$-levels of the
unfilled $3d$ shell of transition metal ions. The intraatomic
exchange interaction leads to the Hund rule which governs the
occupation of the deep levels. As a result the transition atom in
a semiconductor are magnetic. We will show below that the
interplay between the magnetic impurity scattering and attractive
electron-hole interaction in excited 2DEG results in a bound spin
exciton. The spectrum of these bound states is the subject of the
present study.
\section{Model and approximations}
We start with the model of 2DEG doped by magnetic impurities which
is discussed in detail in \cite{DFVK}. This model is described by
the Hamiltonian
\begin{equation}
H=H_b + H_{i}
\label{1.1}
\end{equation}
where
\begin{equation}
H_b = \sum_
{nm\sigma}E_{n\sigma} a^\dagger_{nm,\sigma}a_{nm,\sigma}
\label{1.2}
\end{equation}
is the Hamiltonian of a 2DEG, strongly quantized by a magnetic
field. It is convenient to  use the symmetric cylindrical gauge
for the vector potential ${\bf A} =(-\frac{B}{2}y,\frac{B}{2},0)$,
so the index $m$ describes different orbital states in a given
Landau level
\begin{equation}
E_n=\hbar \omega_B(n+\frac{1}{2})
\equiv\frac{\hbar^2}{2m}\frac{2n+1}{l_b^2} .
\label{1.3}
\end{equation}

Then, assuming a small impurity concentration, the impurity
related part $H_i$ of the Hamiltonian (\ref{1.1}) can be written
in a general form
\begin{equation}
H_i  =  \sum_\Gamma |i\Gamma\rangle E_{\Gamma}\langle i\Gamma| +
\label{1.4} \\
\end{equation}
\[
\sum_{\gamma}\sum_{nm\sigma}\left(
\langle i\gamma |V|nm\rangle
d^\dagger_{i\gamma\sigma}a_{nm\sigma} + H.c.\right) +
\sum_{nm\sigma}
\langle nm|\Delta V|n'm'\rangle
a^\dagger_{nm\sigma}a_{n'm'\sigma}
\]
Here the state of isolated impurity ion is characterized by a
configuration $d^n$ of its unfilled $d$-shell in a crystal field
preserving the point symmetry of bulk semiconductor (we assume
that the potential, responsible for the confinement in the $z$
direction does not disturb the crystalline environment of the
impurity site). Then the electrons in $d$ shell are characterized
by the representations $\gamma=t_2,e$ of the tetrahedral point
group, and the many-electron states $|\Gamma\rangle$ of the $3d$
shell may be represented as $$d^n_\Gamma = \left(
ne_\uparrow^{r_1} e_\downarrow^{r_2}t_{2\uparrow}^{r_3}
t_{2\downarrow}^{r_4} \right)_{\sum_ir_i=n}$$ ($\downarrow$ and
$\uparrow$ are two projections of the electron spin). The
scattering part of $H_i$ consists of two components \cite{KF94}:
the second term in eq. (\ref{1.4}) describes the resonance part of
the impurity scattering. The third one represents the short-range
substitution potential $\Delta V = V_i(r-R_0) - V_{host}(r-R_0)$
where $R_0$ is the position of the substitution impurity in a host
lattice. The resonance scattering arises together with the usual
"potential" scattering due to the fact that the energy level
$\varepsilon_{i\gamma} = E_\Gamma(d^n)-E_{\Gamma'}(d^{n-1})$
enters the fundamental energy gap of the host semiconductor or
appears in the lowest conduction band or the topmost valence band.
Here the configuration $d^{n-1}$ misses one electron in a state
$\gamma=e$ or $t_2$ in comparison with the state $d^n$.

The impurity problem with the two scattering mechanisms can be
solved in the general case \cite{Pic84,FK86}. In our special case
its solutions are essentially different for $e$ and $t_2$
channels. As is shown in \cite{DFVK}, the $t_2$ component of
impurity potential results in deep levels in forbidden energy gap
or resonances in the conduction band, but only weakly disturbs the
Landau levels $E_n$. On the other hand, $e$-scattering results in
an appearance of the bound Landau states with $m=0$ between Landau
levels with nonzero orbital quantum number and in a splitting of
the lowest bound state from the Landau grid. The reason for this
difference is in the {\em short-range} nature of both components
of the scattering potential. The eventual reason for this
difference is the orbital dependence of the matrix element
\begin{equation}
\langle nm|\Delta U|\gamma \mu \rangle \sim \left( \frac{\rho
}{l_{B}} \right) ^{|m|}\ll 1
\label{sel}
\end{equation}
for $m\neq 0$ \cite{DFVK}. Here $\rho$ is the radial variable in
the cylindrical coordinates $(\rho,\varphi,z)$. When estimating
this matrix element one should choose $\rho$ of the order of the
atomic radius. As a result only the orbitals $|e1\rangle \propto
|r^{2}-3z^{2}\rangle \sim Y_{20}$ with $m=0$ may be strongly
hybridized with the Landau states.

According to \cite{DFVK,FK86}, the localized eigenstates of the
Hamiltonian (\ref{1.1}), (\ref{1.4}) are given by the following
equation
\begin{equation}
E_{i\gamma\sigma }-\varepsilon _{i\gamma\sigma }-
M_{\gamma }(E_{i\gamma\sigma })=0,
\label{DM3}
\end{equation}
where
\begin{equation}
M_{\gamma }(E_{i\gamma\sigma}) = \sum_{\beta }\frac{\langle \gamma
\mu |\Delta U|\beta \rangle \langle \beta|\Delta U|\gamma \mu
\rangle } {E_{i\gamma\sigma}-E_{\beta}}.
\label{SE}
\end{equation}
where $|\beta \rangle\equiv|bn\sigma\rangle$ stand for the
eigenfunctions of the magnetically quantized electrons captured by
the local potential $\Delta U$. The latter problem was solved in
ref. \cite{Avishai93} for attractive  potential $\Delta U<0$ . It
was shown that the short range potential also perturbs only the
zero orbital states $m = 0$. The general structure of the electron
spectrum of doped 2DEG is sketched in Fig.1.
\begin{figure}[htb]
\epsfysize=19 \baselineskip
\centerline{\hbox{\epsffile{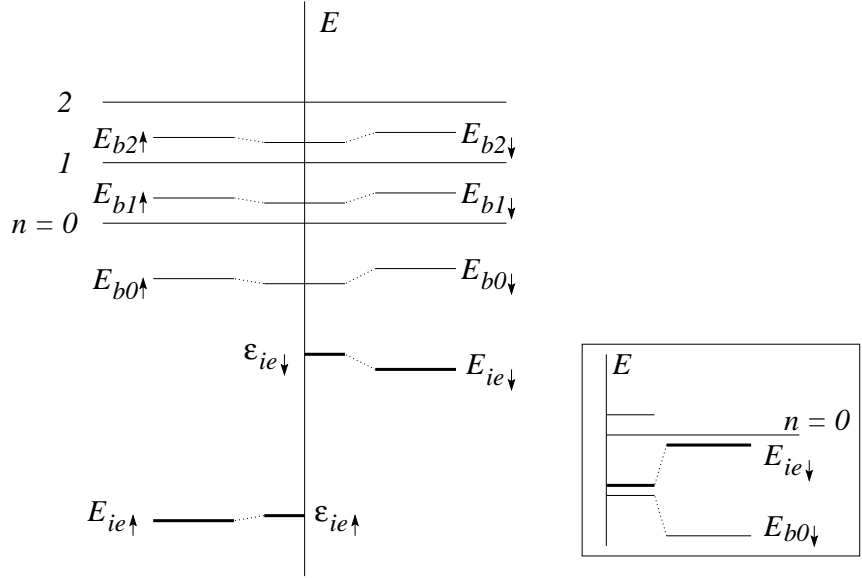}}} \caption{A schematic
representation of interacting Landau and impurity levels.
$E_{\mbox{ie}\sigma}$ are the impurity levels for up- and
down-spin electrons. They result from the prime impurity levels
$\varepsilon_{ie\sigma}$ shifted to new positions due to their
interaction with the Landau levels with $m=0$. The Landau levels
with $m=0$ are also shifted to their new positions $E_{be\sigma}$.
The insert illustrates the case when an impurity prime spin-down
state is nearly degenerate the lowest Landau spin-down state.}
\end{figure}
%


This observation allows us to divide the states in the Landau band
(\ref{1.2}) into two groups,
\begin{equation}
H_b = H_{b0} + H^{\prime}_{b},
\label{1.6}
\end{equation}
Here $H_{b0}$ includes only the states with the zero angular
moment $m=0$, whereas $H^\prime_b$ includes all the remaining
states. Only the states with $m=0$ are involved in the formation
of the bound Landau states given by the solutions of eq.
(\ref{DM3}). Similar reduction can be made in the manifold
$\{i\gamma\sigma\}$ due to the `selection rule' (\ref{sel}): when
considering the renormalization of the states $|\beta\rangle$ due
to the resonance scattering, we leave only the state $|e1\rangle$
in the corresponding sector of the secular matrix.  As a result,
one has the following equation
\begin{equation}
E_{bn\sigma}-E_{bn\sigma}^{(0)} = \frac
{ \langle bn\sigma |\Delta U|e1\sigma\rangle\langle e1\sigma|
\Delta U |bn\sigma \rangle}{E_{bn\sigma}-E_{ie\sigma}} +
\delta M(E_{bn\sigma}).
\label{1.7}
\end{equation}
for the energy shift of a given level $E_{bn\sigma}^{(0)}$ which
according to \cite{Avishai93} appears in the gap between the bare
Landau levels $E_{n-1}$ and $E_n$. Here we have picked up the
direct mutual repulsion of the $n$-th Landau level and the
$d$-level in the first term of the right-hand side of eq.
(\ref{1.7}). The influence of the other bound Landau states is
given by the second term,
$$\delta M(E) = \sum_{\beta'\neq \beta} \frac{ \langle
\beta'|\Delta U|e1\sigma\rangle\langle e1\sigma| \Delta U |\beta'
\rangle(E-E^{(0)}_\beta)} {(E-E_{ie\sigma})(E-E^{(0)}_{\beta'})}.
$$
The role of these states is to keep the renormalized levels in the
same energy interval $E_{n-1,\sigma} < E_{bn\sigma} <
E_{n\sigma}$.

An important feature of the resonant channel of impurity
scattering is its {\em spin selectivity}, which stems from the
spin structure of the transition metal $d$ shell. It is known
\cite{KF94} that the transition metal impurities follow in their
main features the `Aufbau principle' of the quantum mechanics of
isolated atoms. This means that the d-shell is usually filled in
accordance with the Hund's rule or, in other words, the exchange
interaction makes the level $E_{ie\uparrow}$ lie always below the
level $E_{ie\downarrow}$. Then the splitting $\Delta_{es} =
E_{ie\downarrow} - E_{ie\uparrow}$ generates the spin splitting
$\Delta_{bs}^{(n)}$ of the bound Landau states via eq.
(\ref{1.7}). This splitting can be estimated as
\begin{equation}
\Delta^{(n)}_{bs} \approx
\frac{\Delta_{es}}{\Delta^{(n)}_\uparrow\Delta^{(n)}_\downarrow}
\left(
|V_{en}|^2 + \sum_{n'\neq n}\frac
{|V_{en'}|^2 \delta_n}
{E^{(0)}_{bn}-E^{(0)}_{bn'}}
\right)
\label{SSS}
\end{equation}
Here $\Delta^{(n)}_\sigma=E_{bn}-E_{ie\sigma}$, $V_{en}=\langle
e1|\Delta U |bn \rangle$, $\delta_n=E_{bn}-E^{(0)}_{bn}$ (the spin
dependence of two last quantities is neglected). It is important
for the further classification of the spin excitons that the sign
of $\Delta^{(n)}_{bs}$, unlike the sign of $\Delta_{es}$, can be
both negative or positive, since the energy differences
$\Delta^{(n)}_\uparrow$ and $\Delta^{(n)}_\downarrow$ can have
either the same or the opposite signs depending on the type of the
transition metal ion and the host matrix (see discussion in
\cite{DFVK}). According to eq. (\ref{SSS}) the resonance impurity
scattering results in a `transfer' of the exchange splitting
$\Delta_{es}$ in the impurity $d$-shell to the spectrum of Landau
electrons. This transfer is illustrated by the level
renormalization in Fig. 1.

Now we know the general structure of the one-particle spectrum of
a magnetically doped 2DEG, which should serve as a background when
magnetic excitons and magnetoplasmons are formed. This spectrum
consists of equidistant Landau levels with $m\neq 0$ which are
unperturbed by the impurity scattering. The states with $m=0$ form
their own grid: a pair of spin split states $E_{bn\sigma}$ appears
in each energy gap $E^{(0)}_{n}-E^{(0)}_{n-1}$, and the lowest
pair of bound states $E_{bn\sigma}$ arises in the fundamental
energy gap of the 2D semiconductor below the Landau level $E_0$.
\section{Energy spectrum of localized spin excitons}
According to the general classification of the multiparticle
excitations from a filled Landau level in 2DEG \cite{KH84}, the
lowest branch of the electron-hole excitations is that of spin
waves. These spin excitons arise as bound electron-hole pairs with
parallel spins as a result of a spin-flip excitation from the
lower filled sublevel $E_{0\uparrow}$ of the lowest Landau level
to its higher empty sublevel $E_{0\downarrow}$. To calculate the
spectrum of the {\em localized} spin excitons, one should keep
only the states with $n=0$ in the bare Hamiltonian $H_b$
(\ref{1.2}), take into account the Zeeman splitting of the Landau
states explicitly and add the electron-hole Coulomb interaction
$H_{int}$,
\begin{equation}
H^0_{ex}=\sum_p (E_{0\uparrow} a^+_pa_p +
E_{0\downarrow} b^+_pb_p)
+ H_{int}.
\label{ham2}
\end{equation}
Operators $a^+(a)$ and $b^+(b)$ describe creation (annihilation)
of electrons with the up or down spins $\sigma$, index $n=0$ being
omitted. Here the electron wave functions are written in the
asymmetric Landau gauge ${\bf A}= B(y,0,0)$. Then the total
Hamiltonian of a doped 2DEG takes the form
\begin{equation}
H=H^0_{ex}+H_{i}
\label{ham1}
\end{equation}
The impurity related term $H_i$ was analyzed in the previous
section using the symmetric gauge. Now it should be re-derived in
the Landau gauge.

Using the solution of the impurity problem found in the previous
section and taking into account that the Landau levels with $m=0$
are included in $H_{ex}$ contrary to `extraction' principle
formulated in eq. (\ref{1.6}), we can write $H_i$ in the form (see
Appendix),
\begin{equation}
H_i = (|A_{00}|^2 E_{b0\uparrow} -  E_{0\uparrow}) a^\dagger_{i0}
a_{i0} + (|B_{00}|^2 {E}_{b0\downarrow} - E_{0\downarrow})
b^\dagger_{i0} b_{i0} + H_{i,nd}. \label{ham3}
\end{equation}
Here $E_{b0\sigma}$ are the solutions of eq. (\ref{1.7}) for the
lowest bound Landau state, and the operators $a^\dagger_{i0}$ and
$b^\dagger_{i0}$ create the corresponding eigenfunctions in the
symmetric cylindrical gauge. The coefficients $A_{00}$, $B_{00}$
are defined in eq. (\ref{A.02}). The off-diagonal term  $H_{i,nd}$
contains the contribution from the higher Landau levels with $n>0$
whose value is of the order of $\sim |A_{0n}|^2/E_n \ll 1$ and
will be neglected in the further calculations.

In principle, the scattering potential contains also the terms
corresponding to spin-flip processes,
\begin{equation}
H_{\perp}=J(
d^\dagger_{ie\uparrow}d_{ie\downarrow}b^\dagger_{i0}a_{i0} +
d^\dagger_{ie\downarrow}d_{ie\uparrow}a^\dagger_{i0}b_{i0}).
\label{spinflip} \end{equation}
The spin-flip terms are inessential in comparison with the leading
spin-diagonal terms (\ref{ham3}) because each spin flip process
costs the Hund energy $\Delta_{es} = E_{ie\downarrow} -
E_{ie\uparrow}$, so that $J\sim |V_{eo}|^2/\Delta_{ex}$. The
spin-flip processes, in principle result in multiple creation of
spin excitons, but the contribution of these processes to the
spin-wave spectrum is at least $\sim J^2$, and we also neglect
them in the further calculations.

Then, we turn from the cylindrical gauge to the Landau gauge,
\begin{equation}
a^\dagger_{i0}=\sum_p A_p a^\dagger_p,~~~
b^\dagger_{i0}=\sum_p B_p b^\dagger_p,
\label{2.0}
\end{equation}
where
\begin{equation}
A_p=\langle i0\uparrow | p\uparrow\rangle,~~~
B_p=\langle i0\downarrow | p\downarrow\rangle
\label{A.1}
\end{equation}

As a result the impurity Hamiltonian in the Landau gauge has the
form
\begin{equation}
H_i =\sum_{pp'}\left[ U_\uparrow(p,p') a^\dagger_{p}a_{p'} +
U_\downarrow(p,p') b^\dagger_{p}b_{p'}
\right]
\label{2.1}
\end{equation}
where
\begin{equation}
U_\uparrow(p,p')=K_{\uparrow}J_{p\uparrow}J^*_{p'\uparrow},~~
U_\downarrow(p,p') = K_{\downarrow} J_{p\downarrow}
P^*_{p'\downarrow}
\label{2.2}
\end{equation}
Here where $J_{p\sigma} = \langle \psi_{b\sigma} |\psi_{p,0
\sigma} \rangle$, and $\psi_{p,0\sigma}$ are the wave functions of
the lowest Landau level,
\begin{equation}
\psi_{p,0\sigma}(x,y)=\frac{1}
{(2\pi^{3/2})^{1/2}}e^{ipy}
e^{-\frac{(x+p)^2}{2}}~.
\label{A.4}
\end{equation}
Here $x,\ y$ are Cartesian projections of the dimensionless vector
${\bf r}/l_B$. The coefficients $K_\sigma$ are calculated in
Appendix.

To derive the impurity Hamiltonian in the Landau gauge, we use the
identity
\begin{eqnarray}
a^\dagger_pa_{p'} =
\int\frac{dq_x}{2\pi}e^{-iq_x\left(\frac{p+p'}{2}\right)}
\rho_\uparrow(q_x,p'-p)\nonumber\\
b^\dagger_pb_{p'} =
\int\frac{dq_x}{2\pi}e^{-iq_x\left(\frac{p+p'}{2}\right)}
\rho_\downarrow(q_x,p'-p),
\label{2.3}
\end{eqnarray}
where the electron density operators are
\begin{equation}
\begin{array}{c}
\rho_\uparrow({\bf q}) = \sum_p a^+_pa_{p+q_y}
e^{iq_x(p+\frac{q_y}{2})}\\
\rho_\downarrow({\bf q}) = \sum_p b^+_pb_{p+q_y}
e^{iq_x(p+\frac{q_y}{2})}
\end{array}
\end{equation}
with ${\bf q}=(q_x,q_y)$. Then inserting (\ref{2.3}) into eq.
(\ref{2.1}), we get after straightforward calculations
\begin{equation}
H_{i}= \sum_{{\bf q},\sigma}U_\sigma({\bf q})
e^{-\frac{q^2}{4}}\rho_\sigma({\bf q})
\label{ham4}
\end{equation}
where the matrix elements of the impurity potential are
\begin{equation}
U_\sigma({\bf q}) = K_\sigma\int d^2r_1 d^2r_2
\psi_{b\sigma}\left(\frac{r^2_1}{2}\right) \psi_{b\sigma}
\left(\frac{r^2_2}{2} \right)e^{-\frac{|{\bf r}_1-{\bf
r}_2|^2}{4}}\times \label{2.4}
\end{equation}
$\exp \{\frac{1}{2}\left[(x_1-x_2)q_y + (y_1-y_2)q_x
-i(y_1+y_2)q_y + i(x_1+x_2)q_x - i(x_1+x_2)(y_1-y_2)\right]\}
$
\medskip\\
Here the coefficients $K_\sigma$ determined in Appendix depend on
the specific form of the scattering impurity potential acting on
the electrons in the Landau subband $0\sigma$ [see eqs.
(\ref{A.1}), (\ref{A.2}), (\ref{A.3}), (\ref{A.6}), (\ref{A.7})].
We discuss here the limit of a strong magnetic field when the
cyclotron energy $\hbar\omega_B=\hbar^2/2ml_B$ is large compared
to the Coulomb energy $e^2/\kappa l_B$. It is essential that the
impurity potential is spin selective, i.e., its  components acting
on the electrons in the two Landau subbands can differ
significantly in magnitude (see below).

We consider the case of the filling factor $\nu=1$, when the
spin-up Landau band is totally full and the spin-down Landau band
is completely empty. Then the eigenfunctions of the Hamiltonian
(\ref{ham2})
\begin{equation}
\Psi^{(0)}_{ex,{\bf k}} = \sum_p b^+_pa_{p+k_y}
e^{ik_x(p+\frac{k_y}{2})}|0\rangle
\label{wavefun1}
\end{equation}
correspond to free spin-excitons with the energy spectrum
$\varepsilon_{ex}(k)$,
\begin{equation}
H^{(0)} \Psi^{(0)}_{ex}(k) = \varepsilon_{ex}(k)
\Psi^{(0)}_{ex,{\bf k}}.
\label{ham5}
\end{equation}
Here ${\bf k}=k_x,k_y$ is the wave vector of a spin exciton. The
exciton dispersion law is
\begin{equation}
\varepsilon_{ex}(k)=\Delta_B +\left(\frac{e^2}{\kappa l_B}\right)
\left(\frac{\pi}{2}\right)^{1/2}[1-e^{-k^2/4}I_0(k^2/4)]
\equiv \Delta_B+\Omega(k^2)
\label{disp}
\end{equation}
(see \cite{Lelo80,Bychok81,KH84,Bychok94}). Here $\Delta_B =
|g\mu_B B|$ is the Zeeman energy, $I_0$ is a modified Bessel
function.

The wave function of a bound exciton is looked for in the form
\begin{equation}
\Psi_{ex} = \sum_{\bf k} f({\bf k})\Psi^{(0)}_{ex,{\bf k}}
\label{wavefun2}
\end{equation}
The function (\ref{wavefun2}) must be an eigenfunction of the
Hamiltonian (\ref{ham1}). The standard procedure leads to the
equation
$$[\varepsilon_{ex}(k) - \varepsilon] f({\bf k}) +$$
\begin{equation}
2i\sum_{\bf k'}\tilde U_+({\bf k} - {\bf k'}) \sin\frac{1}{2}
[{\bf k'} \times {\bf k}]_z f({\bf k'}) - 2\sum_{\bf k'} \tilde
U_-({\bf k} - {\bf k'}) \cos \frac{1}{2} [{\bf k'} \times {\bf
k}]_z f({\bf k'}) = 0, \label{ham6}
\end{equation}
for the envelope function $f({\bf k})$. Here
$$\tilde U_\pm({\bf q}) = \frac{1}{2}(U_{\uparrow}({\bf q})\pm
U_{\downarrow}({\bf q}))e^{-\frac{q^2}{4}}\equiv U_\pm({\bf
q})e^{-\frac{q^2}{4}} $$

As is discussed in Appendix, the localization radius $\rho_b$ of
the impurity wave function is essentially smaller than the
magnetic length. Then the $q$-dependence of the matrix elements
(\ref{2.4}) is insignificant, and they can be estimated as
\begin{equation}
U_\sigma \approx K_\sigma(2\pi^2)\left( \int_0^\infty
d\xi\psi_{b\sigma}(\xi)\right)^2 \equiv K_\sigma I_b \label{2.5}
\end{equation}
Taking into account the cylindrical symmetry of the problem, the
solutions of equation (\ref{ham6}) are looked for in the form
\begin{equation}
f({\bf k}) = f_m(k)e^{im\varphi}
\label{wavefun3}
\end{equation}
where the integer quantity $m$ is the quantum number of the bound
exciton.

Now we substitute the functions (\ref{wavefun3}) in eq.
(\ref{ham6}) and carry out the integration over the directions of
the vector ${\bf k}'$. Then the term proportional to $U_+$
contains the integrals
\begin{equation}\label{uminus}
\int_0^{2\pi}\frac{d\varphi}{2\pi}e^{\frac{kk'}{2}\cos\varphi}
\sin(\frac{kk'}{2}\sin\varphi)\cdot\sin m\varphi =
\frac{(kk')^{|m|}}{2^{|m|+1}|m|}\mbox{sign}(m)
\end{equation}
in which the sign function is defined as
\begin{equation}\label{theta}
\mbox{sign}(m) = \left\{\begin{array}{cc}
 1, & \mbox{if}\ m > 0, \\
 0, & \mbox{if}\ m = 0, \\
  -1, & \mbox{if}\ m < 0.
\end{array}\right.
\end{equation}
The term proportional to $U_-$ contains the integrals
\begin{equation}\label{uplus}
\int_0^{2\pi}\frac{d\varphi}{2\pi}e^{\frac{kk'}{2}\cos\varphi}
\sin(\frac{kk'}{2}\sin\varphi)\cdot\cos m\varphi = \displaystyle
  \frac{(kk')^{|m|}}{2^{|m|+1}|m|}
\end{equation}
for all values of the quantum number $m$.

The equations defining the radial parts $f_m(k)$ of the envelope
functions (\ref{wavefun3}) are
\begin{equation}
f_m(k) = W_m k^{|m|}e^{-\frac{k^2}{4}} \frac{1}{\varepsilon -
\varepsilon_{ex}({k})} F_m \label{wavefun40}
\end{equation}
where
$$ W_m = \frac{1}{2^{|m|+1}|m|!}\{U_\downarrow [1 +
\mbox{sign}(m)] - U_\uparrow[1 - \mbox{sign}(m)]\} $$
and
$$F_m = \sum_{\bf k} e^{-\frac{k^2}{4}}k^{|m|} f_m(k).$$

The energy $\varepsilon_m$ of the bound exciton with the quantum
number $m$ can be found as a solution of the equation
\begin{equation}\label{secular}
1 = W_m M_m(\varepsilon_m),
\end{equation}
where
\begin{equation}
\label{M1} M_m(\varepsilon) = \sum_{\bf k} \displaystyle
\frac{k^{2|m|}e^{-\frac{k^2}{2}}}{\varepsilon -
\varepsilon_{ex}({k})}.
\end{equation}

We introduce the new variable $\omega = k^2$ and convert summation
in equation (\ref{M1}) in integration. Then the fact that at small
$\omega$ the dispersion law $\Omega(\omega)\propto \omega$ allows
us to find the behavior of the quantity $M_m(\varepsilon)$ at
$\varepsilon \rightarrow \Delta_B$.
$$M_m(\varepsilon) = \pi\int_0^D d\omega\frac{\omega^m
e^{-\frac{\omega}{2}}} {\varepsilon - \Delta_B - \Omega(\omega)}
\approx$$
\begin{equation}
\label{M2}
 \left\{
\begin{array}{ll}
-2\pi m_{ex}^*\ln m^*_{ex}|\Delta_B-\varepsilon|, & m=0 \\ & \\
M_m(\Delta_B) - M^\prime_m(\Delta_B)|\varepsilon - \Delta_B|, &
\varepsilon \to \Delta_B-0,~ m\neq 0.
\end{array}
\right.
\end{equation}
Here $m*_{ex}=\displaystyle \frac{2\kappa\hbar^2}{e^2} \left(
\frac{2eB}{\pi\hbar c} \right)^{1/2}$ is the effective mass of the
free exciton at small momenta \cite{Lelo80}, $M_m(\Delta_B)<0,$~
$M'(\Delta_b) = - \left. \displaystyle
\frac{dM(\varepsilon)}{d\varepsilon}
\right|_{\varepsilon-\Delta_B\to -0}>0$.

Now using the normalization condition
$$\sum_{\bf k}|f_m(k)|^2 = 1$$
one finds that
\begin{equation}
F_m =|W_m|\left[ M'(\varepsilon_m)\right]^{-1/2} . \label{self}
\end{equation}
This equation closes the procedure. Now the spectrum of spin
excitations in 2DEG pinned by the resonance impurity with its own
localized spin is determined by eqs. (\ref{secular}) - (\ref{M2}).
The coupling constants are determined in eqs. (\ref{2.5}),
(\ref{A.6}), (\ref{A.7}), and the wave function of the bound spin
exciton is given by eqs. (\ref{wavefun2}), (\ref{wavefun40}),
(\ref{self}).

Starting the analysis of the bound exciton solutions with the case
of $m=0$, we see that in the weak scattering limit the magnitudes
of both coupling constants (\ref{2.5}) are determined by the
coefficient (\ref{A.6}). Then the potential in eq. (\ref{secular})
is given by the following equation
\begin{equation}
W_0 = -U_-= - V_{eb}I_b(\beta_\uparrow -\beta_\downarrow)=
V^2_{eb}I_b \frac{\Delta_{es}}{\Delta_\uparrow\Delta_\downarrow}
> 0. \label{weak}
\end{equation}
The scattering potential (\ref{weak}) is repulsive and the eq.
(\ref{secular}) for $m = 0$ has no bound solutions below the
exciton band.

Such a solution appears when the resonance scattering for the spin
down electrons is strong (see insert in Fig.1). This is a
realistic condition for transition metal impurities \cite{KF94}.
In that case we use eq. (\ref{A.6}), for the potential acting on
the spin up electrons, and eq. (\ref{A.7}), for the potential
acting on the spin down electrons. The coupling constant is
\begin{equation}
W_0 = - U_-\approx - \frac{1}{2}(V_{eb}-w_0)I_b \label{strong}
\end{equation}
Due to the logarithmic divergence of the function
$M_0(\varepsilon)$ near the bottom of the excitonic band (see eq.
(\ref{M2})), the discrete solution of eq. (\ref{secular}) appears,
provided the resonance component of impurity scattering is
stronger than the potential one, $V_{eb}>w_0$, i.e, $W_0 < 0$ (see
Fig.2).
\begin{figure}[htb]
\epsfysize=19 \baselineskip
\centerline{\hbox{\epsffile{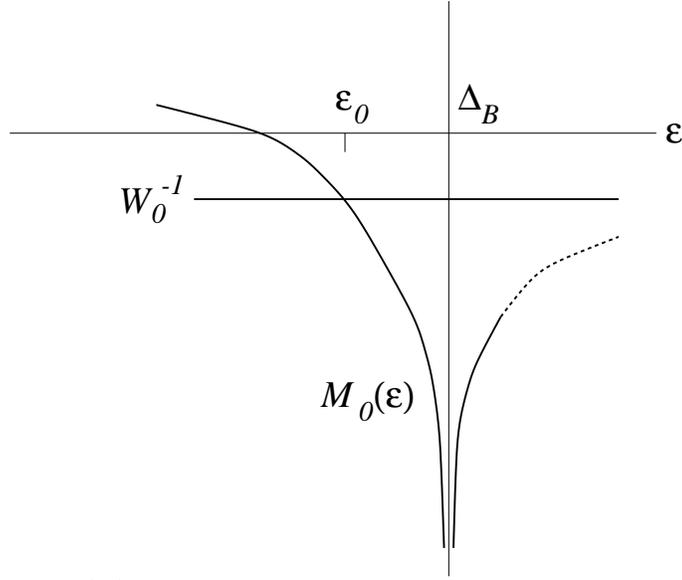}}} \caption{A graphical
solution of eq. (\ref{secular}) for $m=0$. Due to the logarithmic
divergence of the function $M_0(\varepsilon)$ at $\varepsilon =
\Delta_B$, a spin-exciton can be always bound with an energy
$\varepsilon_0$ below the bottom of the spin-exciton
band.}\end{figure}
\begin{figure}[htb]
\epsfysize=12 \baselineskip
\centerline{\hbox{\epsffile{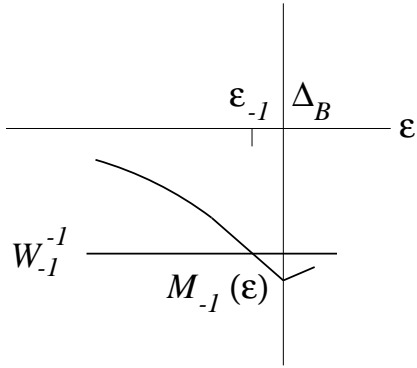}}} \caption{A graphical
solution of eq. (\ref{secular}) for $m=-1$. The function
$M_{-1}(\varepsilon)$ is now non-divergent. Hence, a spin-exciton
can be bound with an energy $\varepsilon_{-1}$ below the bottom of
the spin-exciton band if the parameter $W_{-1}$ is larger than the
threshold value $\overline{W}_{-1} =
1/M_{-1}(\Delta_B)$.}\end{figure}
%

The selectivity with respect to the orbital quantum number $m$ of
the bound spin exciton is intrinsically connected with the spin
selectivity of the impurity potential. This orbital momentum is
the sum of the electron and hole momenta, $m=m_e+m_h$. As was
shown previously \cite{DFVK,Avishai93}, only the electrons or
holes with $m_{e,h}=0$ can be captured by the short-range impurity
potential which is $U_\downarrow$ for electrons and $-U_\uparrow$
for holes. Therefore, in a bound exciton with $m\neq 0$ one of the
carriers (electron or hole) must have zero momentum. Then the
momentum of the whole exciton is, in fact, the momentum of the
second carrier. This second carrier is bound by the combined
action of the Coulomb attraction of the first carrier and the
diamagnetic contribution of the Lorenz force. The sign of the
orbital momentum and the charge of the carrier predetermines the
sign, attractive or repulsive, of Lorenz force contribution to the
total potential. As a result, only the electrons with $m_e>0$ and
the holes with $m_h<0$ can be captured in the limit of a strong
magnetic field. Hence, the sign of the exciton orbital momentum
provides an information on which carrier is bound by the
short-range potential. It is the hole, bound by the potential $-
U_\uparrow$, in the case of $m>0$ and the electron, bound by the
potential $- U_\uparrow$, in the case of $m<0$.

In the case of $m=0$ both carriers are captured by their
corresponding short-range potentials and then their total coupling
strength, $U_\downarrow - U_\uparrow$, determines the binding
energy of the exciton.

\section{Conclusions}

To conclude, we have found that the magnetic impurity can bind the
spin exciton in a 2DEG. It turned out that the mechanism of the
exciton capture is the {\em spin-selective resonance scattering}
by the deep impurity levels. This spin selectivity is stems from
the Hund rule in particular case of transition metal impurities.
The interaction in the second order in scattering potential can be
described in terms of an indirect spin exchange $J$ (see eq.
\ref{spinflip}), and only the longitudinal component of this
exchange is essential for the formation of a bound spin exciton.
The transversal components of this exchange give the contribution
to binding energy only in 4th order in the scattering potential.

It is found that the spin-selective impurity potential is always
capable of binding the exciton in a state with the moment $m=0$
because of the threshold van Hove singularity of the density of
states in a 2DEG. Excitons with $m\neq 0$ can be also trapped, but
then the conditions for the capture are more severe.

The analysis of the electronic structure of 3d transition metal
impurities in GaAs-related systems shows that the strongest
binding potential is created by the light elements (V, Cr).

\section{Acknowledgements}

This work was supported by the German-Israeli Foundation for
Research and Development, Grant No.0456-220.07195. K.K. thanks
Israeli Science Foundation for support (grant ''Nonlinear Current
Response of Multilevel Quantum Systems''). The Authors are
indebted to Yu. Bychkov, S. Dickmann, T. Maniv and I. Vagner for
valuable discussions.

%

\vspace{1cm}
\appendix{\bf APPENDIX}
\vspace{1cm}

\setcounter{equation}{0}
\renewcommand{\theequation}{A.\arabic{equation}}

In order to derive the impurity Hamiltonian for 2D electrons in
the Landau gauge, we use the fact that the resonance scattering
involves only the Landau states with $m=0$. These states are
included in the Hamiltonian (\ref{ham2}), so one should subtract
them from $H_i$, which then acquires the form
\begin{equation}
H_i = E_{b0\uparrow}a^\dagger_{i0}a_{i0} +
E_{b0\downarrow}b^\dagger_{i0}b_{i0} -
E_{0\uparrow}a^\dagger_{0}a_{0} -
E_{b0\downarrow}b^\dagger_{0}b_{0} \label{A.01}
\end{equation}
Then the bound Landau states with $m=0$ can be re-expanded in free
Landau states,
\begin{equation} a_{i0}=\sum_n A_{0n}a_{n0},\ \
b_{i0}=\sum_n B_{0n}b_{n0} \label{A.02}
\end{equation}
\cite{Avishai93}. As a result we come to eq. (\ref{ham3}).

In order to calculate the expansion coefficients $A_p$, $B_p$ in
eq. (\ref{2.0}), one needs the wave functions of the bound
electron in the lowest Landau level. The wavefunctions of the
lowest localized Landau states correspond to the solutions
$E_{b0\sigma}$ of eq. (\ref{1.7}). If these states are deep enough
below the unperturbed Landau spectrum, one can neglect the
contribution $\delta M(E)$ of higher Landau levels, and the
eigenfunctions $\psi_{i0\sigma}$ has the form \cite{DFVK}
\begin{equation}
\psi_{i0\sigma}  = - \sin \theta_\sigma \psi_{ie\sigma} +
\cos \theta_\sigma\psi_{b\sigma}
\label{A.2}
\end{equation}
with the mixing coefficient given by $ \tan 2\theta_\sigma = 2
V_{eb}/ \Delta_\sigma$, and $V_{eb} = \langle e1|\Delta
U|b0\rangle$. The wavefunctions $\psi_{b\sigma}$ describe the
Landau state bound in a short range attractive potential
\cite{Avishai93},
\begin{equation}
\psi_{b\sigma}(\xi) =\frac
{\Gamma(\xi)}{l_B\sqrt{2\pi\psi'(\xi)}}
\frac{
W_{\alpha,0}(\xi)}{\xi^{1/2}}, \;\;\;
\label{BWF2}
\end{equation}
where $2\xi=(\rho/l_B)^2$, $\Gamma$, $\psi'$ and $W_{\alpha,m}$
are gamma function, trigamma function and Whittaker function,
respectively. Index $\alpha$ is determined by the corresponding
eigenstate, $\alpha=2^{-1}(1-\epsilon_{b0\sigma}l_B^2)$ with
$\epsilon=2m^*E/\hbar^2$. When the level $\epsilon_{b0\sigma}$ is
deep enough or the magnetic field is weak enough, i.e.
$|\epsilon_{bo\sigma}|l_B^2\gg 1$, the wave function
$\psi_{b\sigma}$ (\ref{BWF2}) has the standard asymptotic,
\begin{equation}
\psi_{b\sigma} \sim \frac{e^{-\varrho}}{\sqrt{\varrho}}
\label{BWF3}
\end{equation}
Here $|\alpha|\approx l_b^2/2\rho_b^2$, $\varrho=\rho/\rho_b$ and
$\rho_b^{-2}=\varepsilon_b$. Thus, the localization radii of the
d-electron $(\rho_d)$, bound Landau electron $(\rho_b)$ and free
Landau electron $(l_B)$ obey the hierarchy $\rho_d \ll\rho_b \ll
l_B$. As a result one can safely neglect the contribution of the
d-component $\psi_{ie\sigma}$ (\ref{A.2}) in the overlap integral
(\ref{A.1}). Then in the case of a week scattering, $\beta_\sigma
= V_{eb}/\Delta_\sigma \ll 1$, $(\theta\ll 1)$, and the overlap
integrals (\ref{A.1}) can be approximated by the following
equations
\begin{equation}
A_{p}=\left(1-\beta_{\uparrow}^2\right)
J_{p\uparrow},~~~
B_{p}=\left(1-\beta_{\downarrow}^2\right)J_{p\downarrow}
\label{A.3}
\end{equation}

Thus the spin dependence of the matrix elements (\ref{A.3}) is
determined by the energy differences $\Delta_\sigma$ in front of
the integral and by the index $\alpha$ of the Whittaker function.
Having in mind the difference in localization degrees of the wave
functions (\ref{BWF2}) and (\ref{A.4}), the magnitude of the
overlap integrals $J_{p\sigma}$ can be estimated as $J_{p\sigma}
\sim (\rho_{b\sigma}/l_B)$, i.e., these integrals are sensitive
both to the spin splitting and to the magnetic field. The energy
differences in the Hamiltonian (\ref{ham3}) are determined by the
short-range component of the impurity potential which in our
theory enters as a phenomenological parameter $w_0$. In a weak
scattering limit $E_{b0\sigma}-E_0=-w_0+\beta_\sigma V_{eb}$ (if
the short-range potential is attractive) \cite{DFVK}. As a result
the latter factor is dominant in this limit, so that
\begin{equation}
K_{\sigma} = -(w_0-\beta_\sigma V_{eb})
\label{A.6}
\end{equation}
In accordance with the Hund's rule for 3d - impurities
$\Delta_\uparrow > \Delta_\downarrow$, so $K_\uparrow >
K_\downarrow$.

Next we consider the situation where one of the resonance $d$ -
levels, namely $\varepsilon_{ie\downarrow}$, is above the lowest
Landau level, and the scattering is strong for the down-spin
electrons, $\theta_\downarrow \sim -\pi/4$. The inequality
$\theta_\uparrow \ll 1$ still holds, since the estimates
(\ref{A.3}) and (\ref{A.6}) for up spin states are still valid. As
for the down spin states, the coefficient $\sin \theta_
{\downarrow}$ is $\sim 1/2$, and
\begin{equation}
K_{\downarrow}=-\frac{1}{2}(w_o+V_{eb})
\label{A.7}
\end{equation}
The magnitudes of the scattering potentials $U_\uparrow$ and
$U_\downarrow$ differ noticeably in this case.

\end{document}